\documentclass[runningheads]{llncs}
\usepackage{graphicx}
\usepackage{amsfonts,amssymb}
\usepackage{bbm}
\usepackage{color}[monochrome]
\usepackage{upgreek}
\usepackage{bm}
\usepackage{adjustbox}
\usepackage[misc]{ifsym}
\newcommand\blfootnote[1]{%
  \begingroup
  \renewcommand\thefootnote{}\footnote{#1}%
  \addtocounter{footnote}{-1}%
  \endgroup
}
\begin{document}
\title{Simultaneous Alignment and Surface Regression Using Hybrid 2D-3D Networks for 3D Coherent Layer Segmentation of Retina OCT Images}
\titlerunning{Hybrid 2D-3D Networks for Alignment and Segmentation of OCT Images}
%
%
\author{Hong Liu\inst{1,2} \and
Dong Wei\inst{2} \and
Donghuan Lu\inst{2} \and
Yuexiang Li\inst{2} \and
Kai Ma\inst{2} \and\\
Liansheng Wang\inst{1}\textsuperscript{(\Letter)} \and
Yefeng Zheng\inst{2}}
\authorrunning{H. Liu et al.}
%
%
\institute{Xiamen University, Xiamen, China\\
\email{liuhong@stu.xmu.edu.cn}, \email{lswang@xmu.edu.cn} \and
Tencent Jarvis Lab, Shenzhen, China\\
\email{\{donwei,caleblu,vicyxli,kylekma,yefengzheng\}@tencent.com}}
\maketitle              
\begin{abstract}
Automated surface segmentation of retinal layer is important and challenging in analyzing optical coherence tomography (OCT).\blfootnote{H. Liu, D. Wei and D. Lu---First three authors contributed equally.}
Recently, many deep learning based methods have been developed for this task and yield remarkable performance.
However, due to large spatial gap and potential mismatch between the B-scans of OCT data, all of them are based on 2D segmentation of individual B-scans, which may loss the continuity information across the B-scans.
In addition, 3D surface of the retina layers can provide more diagnostic information, which is crucial in quantitative image analysis.
In this study, a novel framework based on hybrid 2D-3D convolutional neural networks (CNNs) is proposed to obtain continuous 3D retinal layer surfaces from OCT.
The 2D features of individual B-scans are extracted by an encoder consisting of 2D convolutions.
These 2D features are then used to produce the alignment displacement field and layer segmentation by two 3D decoders, which are coupled via a spatial transformer module.
The entire framework is trained end-to-end.
To the best of our knowledge, this is the first study that attempts 3D retinal layer segmentation in volumetric OCT images based on CNNs.
Experiments on a publicly available dataset show that our framework achieves superior results to state-of-the-art 2D methods in terms of both layer segmentation accuracy and cross-B-scan 3D continuity, thus offering more clinical values than previous works.

\keywords{Optical coherence tomography \and 3D coherent layer segmentation \and B-scan alignment \and 2D-3D hybrid network.}
\end{abstract}
\section{Introduction}
Optical coherence tomography (OCT)---a non-invasive imaging technique based on the principle of low-coherence interferometry---can acquire 3D cross-section images of human tissue at micron resolutions \cite{huang1991optical}.
Due to its micron-level axial resolution, non-invasiveness, and fast speed, OCT is commonly used in eye clinics for diagnosis and management of retinal diseases \cite{abramoff2010retinal}.
Notably, OCT provides a unique capability to directly visualize the stratified structure of the retina of cell layers, whose statuses are biomarkers of presence/severity/prognosis for a variety of retinal and neurodegenerative diseases, including age-related macular degeneration \cite{keane2009evaluation}, diabetic retinopathy \cite{bavinger2016effects}, glaucoma \cite{kansal2018optical}, Alzheimer’s disease \cite{knoll2016retinal}, and multiple sclerosis \cite{saidha2011primary}.
Usually, layer segmentation is the first step in quantitative analysis of retinal OCT images, yet can be considerably labor-intensive, time-consuming, and subjective if done manually.
Therefore, computerized tools for automated, prompt, objective, and accurate retinal layer segmentation in OCT images is desired by both clinicians and researchers.

Automated layer segmentation in retinal OCT images has been well explored.
Earlier explorations included graph based \cite{antony2013combined,garvin2009automated,lang2013retinal,shah2019optimal},
contour modeling \cite{carass2014multiple,novosel2017joint,yazdanpanah2009intra}, and machine learning \cite{antony2013combined,lang2013retinal} methods.
Although greatly advanced the field, most of these classical methods relied on empirical rules and/or hand-crafted features which may be difficult to generalize.
Motivated by the success of deep convolutional neural networks (CNNs) in various medical image analysis tasks~\cite{ker2017deep}, researchers also implemented CNNs
for retinal layer segmentation in OCT images and achieved superior performance to classical methods \cite{he2019fully}. 
However, most previous methods (both classical and CNNs) segmented each OCT slice (called a B-scan) separately given the relatively big inter-B-scan distance, 
despite the fact that a modern OCT sequence actually consists of many B-scans covering a volumetric area of the eye \cite{drexler2008state}.
Correspondingly, these methods failed to utilize the anatomical prior that the retinal layers are generally smooth surfaces (instead of independent curves in each B-scan) and may be subject to discontinuity in the segmented layers between adjacent B-scans, potentially affecting volumetric analysis following layer segmentation.
Although some works \cite{antony2013combined,carass2014multiple,chen2018intraretinal,garvin2009automated,lang2013retinal,novosel2017joint} attempted 3D OCT segmentation,
all of them belong to the classical methods that yielded inferior performance to the CNN-based ones, and overlooked the misalignment problem among the B-scans of an OCT volume.
Besides the misalignment problem, to develop a CNN-based method for 3D OCT segmentation there is another obstacle: anisotropy in resolution \cite{shah2018multiple}.
For example, the physical resolutions of the dataset employed in this work are 3.24 $\upmu$m (within A-scan, which is a column in a B-scan image),  6.7 $\upmu$m (cross A-scan), and 67 $\upmu$m (cross B-scan).

In this work, we propose a novel CNN-based 2D-3D hybrid framework for simultaneous B-scan alignment and 3D surface regression for coherent retinal layer segmentation across B-scans in OCT images.
This framework consists of a shared 2D encoder followed by two 3D decoders (the alignment branch and segmentation branch), and a spatial transformer module (STM) inserted to the shortcuts \cite{ronneberger2015u} between the encoder and the segmentation branch.
Given a B-scan volume as input, we employ per B-scan 2D operations for the encoder for two reasons.
First, as suggested by previous studies \cite{wang2020conquering,zhang2019light},
intra-slice feature extraction followed by inter-slice (2.5D or 3D) aggregation is an effective strategy against anisotropic resolution, thus we propose a similar 2D-3D hybrid structure for the anisotropic OCT data.
Second, the B-scans in the input volume are subject to misalignment, thus 3D operations across B-scans prior to proper realignment may be invalid.
Following the encoder, the alignment branch employs 3D operations to aggregate features across B-scans to align them properly.
Then, the resulting displacement field is employed
to align the 2D features at different scales and compose well-aligned 3D features by the STM.
These 3D features are passed to the segmentation branch for 3D surface regression.
Noteworthily, the alignment only insures validity of subsequent 3D operations, but not the cross-B-scan coherence of the regressed layer surfaces.
Hence, we further employ a gradient-based, 3D regulative loss \cite{wei2018three} on the regressed surfaces to encourage smooth surfaces, which is an intrinsic property of many biological layers.
While it is straightforward to implement this loss within our surface regression framework and comes for free (no manual annotation is needed), it proves effective in our experiments.
Lastly, the entire framework is trained end-to-end.

In summary, our contributions are as following.
First, we propose a new framework for simultaneous B-scan alignment and 3D retinal layer segmentation for OCT images.
This framework features a hybrid 2D-3D structure comprising a shared 2D encoder, a 3D alignment branch, a 3D surface regression branch, and an STM to allow for simultaneous alignment and 3D segmentation of the anisotropic OCT data.
Second, we further propose two conceptually straightforward and easy-to-implement regulating losses to encourage the regressed layer surfaces to be coherent---not only within but also across B-scans, and also help align the B-scans.
Third, we conduct thorough experiments to validate our design and demonstrate its superiority over existing methods.

\section{Method}

\subsubsection{Problem Formulation}
Let $\Omega\subset\mathbb{R}^3$, then a 3D OCT volume can be written as a real-valued function $V(x,y,z):\Omega\rightarrow\mathbb{R}$, where the $x$ and $y$ axis are the row and column directions of a B-scan image, and $z$ axis is orthogonal to the B-scan image.
Alternatively, $V$ can be considered as an ordered collection of all its B-scans: $V(b)=\{I_b\}$, where $I_b:\Phi\rightarrow\mathbb{R}$ is the $b$\textsuperscript{th} B-scan image, $\Phi\subset\mathbb{R}^2$, $b\in[1,N_B]$, and $N_B$ is the number of B-scans.
Then, a retinal layer surface can be expressed by $S=\{r_{b,a}\}$, where $a\in[1,N_A]$, $N_A$ is the number of A-scans, and $r_{b,a}$ is the row index indicating the surface location in the $a$\textsuperscript{th} A-scan of the $b$\textsuperscript{th} B-scan.
That is, the surface intersects with each A-scan exactly once, which is a common assumption about macular OCT images (\emph{e.g.}, in \cite{he2019fully}).
The goal of this work is to locate a set of retinal layer surfaces of interest $\{S\}$ in $V$, preferably being smooth, for accurate segmentation of the layers.

\subsubsection{Method Overview}
\begin{figure}[t]
\centering
\includegraphics[width=.9\textwidth,trim=0 2 0 0,clip]{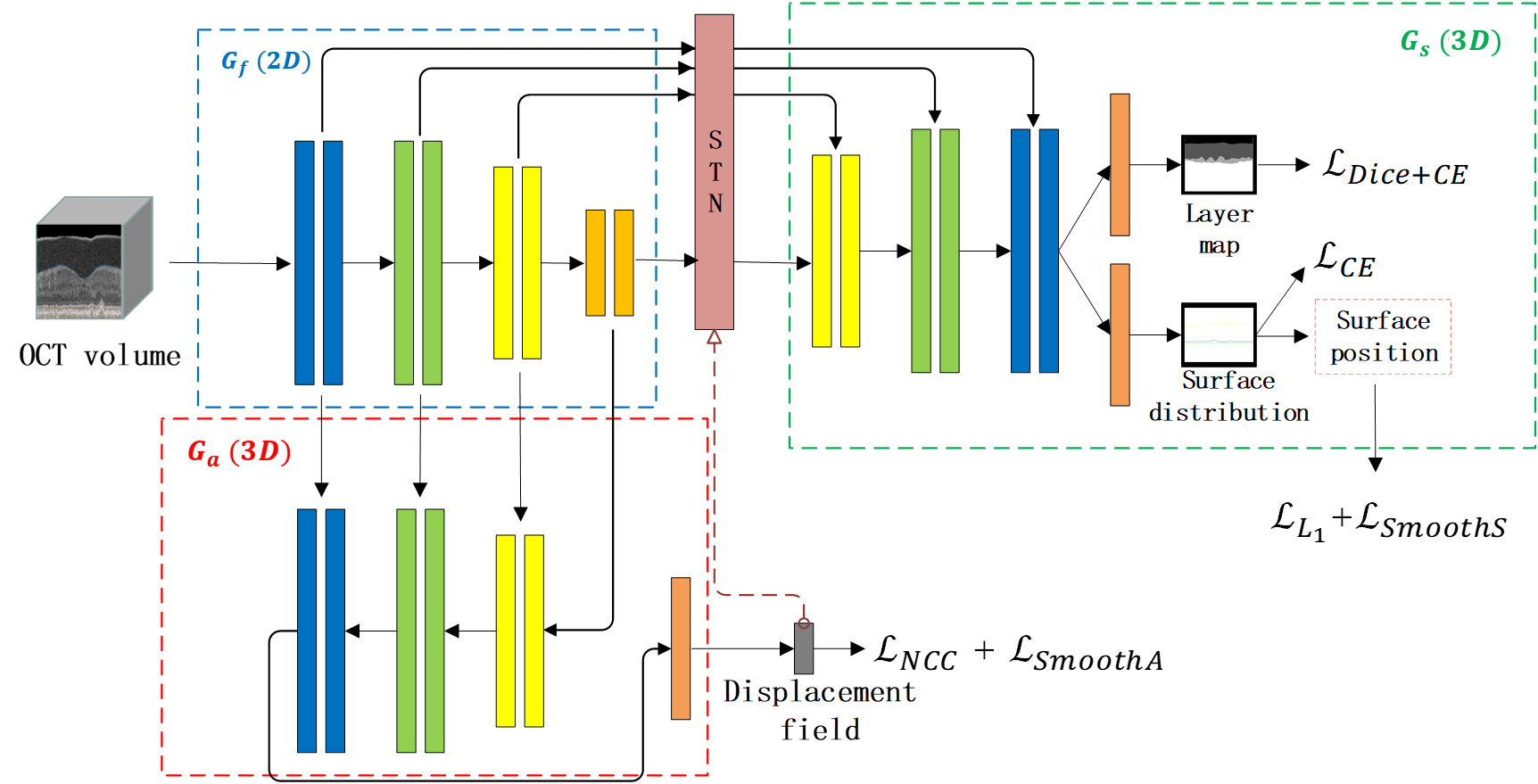}
\caption{Overview of the proposed framework.} \label{fig:overview}
\end{figure}

The overview of our framework is shown in Fig \ref{fig:overview}.
The framework comprises three major components: a contracting path $\mathnormal{G_f}$ (the shared encoder) consisting of 2D CNN layers and two expansive paths consisting of 3D CNN layers $\mathnormal{G_a}$ (the alignment branch) and $\mathnormal{G_s}$ (the segmentation branch), and a functional module: the spatial transformer module (STM).
During feature extraction phase, 2D features of separate B-scans in an OCT volume are extracted by $\mathnormal{G_f}$.
These features are firstly used to generate B-scans alignment displacement by $\mathnormal{G_a}$, which is used in turn to align the 2D features via the STM.
Then, the well-aligned features are fed to $G_s$ to yield final segmentation.
Each of $G_a$ and $G_s$ forms a hybrid 2D-3D residual U-Net \cite{ronneberger2015u} with $G_f$.
The entire framework is trained end-to-end.
As $G_f$ is implemented as a simple adaption to the encoder in \cite{zhou2019models} (3D to 2D),
below we focus on describing our novel $G_a$, $G_s$, and STM.

\subsubsection{B-Scan Alignment Branch}
Due to the image acquisition process wherein each B-scan is acquired separately without a guaranteed global alignment and the inevitable eye movement, consecutive B-scans in an OCT volume may be subject to misalignment \cite{cheng2016motion}.
The mismatch mainly happens along the $y$ axis, and may cause problems for volumetric analysis of the OCT data if unaddressed.
Although it is feasible to add an alignment step while preprocessing, a comprehensive framework that couples the B-scan alignment and layer segmentation would mutually benefit each other (supported by our experimental results), besides being more integrated.
To this end, we introduce a B-scan alignment branch $G_a$ consisting of an expansive path into our framework, which takes 2D features extracted from a set of B-scans by $G_f$ and stacks them along the cross-B-scan direction to form 3D input.
The alignment branch outputs a displacement vector $\triangle\bm{d}$, with each element $d_b$ indicating the displacement of the $b$\textsuperscript{th} B-scan in the $y$ direction.
We use the local normalized cross-correlation (NCC) \cite{balakrishnan2019voxelmorph} of adjacent B-Scans as the optimization objective (denoted by $\mathcal{L}_\mathrm{NCC}$) of $\mathnormal{G_a}$.

As smoothness is one of the intrinsic properties of the retinal layers, if the B-scans are aligned properly, ground truth surface positions of the same layer should be close at nearby locations of adjacent B-scans.
To model this prior,
we propose a supervised loss function to help with the alignment:
\begin{equation}
    \mathcal{L}_\mathrm{SmoothA}={\sum}_{b=1}^{N_B-1}{\sum}_{a=1}^{N_A}
    \big((r^g_{b,a}-d_{b}) - (r^g_{b+1,a}-d_{b+1})\big)^2,
\end{equation}
where $r^g$ is the ground truth.
The final optimization objective of the alignment branch is $\mathcal{L}_\mathrm{Align}=\mathcal{L}_\mathrm{NCC} + \mathcal{L}_\mathrm{SmoothA}$.

\subsubsection{Layer Segmentation Branch}
Our layer segmentation branch substantially extends the fully convolutional boundary regression (FCBR) framework by He \emph{et al.} \cite{he2019fully}.
Above all, we replace the purely 2D FCBR framework by a hybrid 2D-3D framework, to perform 3D surface regression in an OCT volume instead of 2D boundary regression in separate B-scans.
On top of that, we propose a global smoothness guarantee loss to encourage coherent surfaces both within and across B-scans, whereas FCBR only enforces within B-scan smoothness.
Third, our segmentation branch is coupled with the B-scan alignment branch, which boost the performance of each other.

The segmentation branch has two output heads sharing the same decoder: the primary head which outputs the surface position distribution for each A-scan, and the secondary head which outputs pixel-wise semantic labels.
The secondary head is used only to provide an additional task 
for training the network, especially considering its pixel-wise dense supervision.
Eventually the output of the secondary head is ignored during testing.
We follow He \emph{et al.} to use a combined Dice and cross entropy loss \cite{roy2017relaynet} $\mathcal{L}_\mathrm{Dice+CE}$ for training the secondary head, and refer interested readers to \cite{he2019fully} for more details.

\paragraph{Surface Distribution Head}
This primary head generates an independent surface position distribution $q_{b,a}(r|V;\theta)$ for each A-scan, where $\theta$ is the network parameters, and a higher value indicates a higher possibility that the $r$\textsuperscript{th} row is on the surface.
Like in \cite{he2019fully}, a cross entropy loss is used to train the primary head:
\begin{equation}
    \mathcal{L}_\mathrm{CE}=-{\sum}_{b=1}^{N_B}{\sum}_{a=1}^{N_A}{\sum}_{r=1}^{R}
    \mathbbm{1}(r^g_{b,a}=r)\log{ q_{b,a}(r_{b,a}^g|V,\theta)},
\end{equation}
where $R$ is the number of rows in an A scan, $\mathbbm{1}(x)$ is the indicator function where $\mathbbm{1}(x)=1$ if $x$ is evaluated to be true and zero otherwise.
Further, a smooth L1 loss is adopted to directly guide the predicted surface location $\hat{r}$ to be the ground truth:
$
    \mathcal{L}_{L1} = {\sum}_{b=1}^{N_B}{\sum}_{a=1}^{N_A} 0.5t_{b,a}^2\mathbbm{1}(|t_{b,a}|<1)+(|t_{b,a}|-0.5)\mathbbm{1}(|t_{b,a}|\geq1),
$ 
where $t_{b,a}=\hat{r}_{b,a}-r_{b,a}^g$, and $\hat{r}_{b,a}$ is obtained via the soft-argmax:
$
    \hat{r}_{b,a}={\sum}_{r=1}^R rq_{b,a}(r|V,\theta).
$



\paragraph{Global Coherence Loss}
Previous studies have shown the effectiveness of modeling prior knowledge that reflects anatomical properties such as the structural smoothness \cite{wei2018three} in medical image segmentation.
Following this line, we also employ a global smoothness loss to encourage the detected retinal surface $\hat{S}(b,a)=\{\hat{r}_{b,a}\}$ to be coherent both within and across the aligned B-scans based on its gradients:
\begin{equation}
    \mathcal{L}_\mathrm{SmoothS} = {\sum}_{b=1}^{N_B}{\sum}_{a=1}^{N_A}\big\|\triangledown \hat{S}(b,a)\big\|^2.
\label{Grad}
\end{equation}
Finally, the overall optimization objective of the segmentation branch is $\mathcal{L}_\mathrm{Seg}=\mathcal{L}_\mathrm{Dice+CE}+\mathcal{L}_\mathrm{CE}+\mathcal{L}_\mathrm{L1}+\lambda \mathcal{L}_\mathrm{SmoothS}$, where $\lambda$ is a hyperparameter controlling the influence of the global coherence loss.





\subsubsection{Spatial Transformer Module}
The B-Scans displacement field $\triangle\bm{d}$ output by the alignment branch $G_a$ is used to align features extracted by $G_f$, so that the 3D operations of the segmentation branch $G_s$ are valid.
To do so, we propose to add a spatial transformer module (STM) \cite{li2017non} to the shortcuts between $G_f$ and $G_s$.
It is worth noting that the STM adaptively resizes $\triangle\bm{d}$ to suit the size of the features at different scales, and that it allows back prorogation during optimization \cite{li2017non}.
In this way, we couple the B-scan alignment and retinal layer segmentation in our framework for an integrative end-to-end training, which not only simplifies the entire pipeline but also boosts the segmentation performance as validated by our experiments.


\section{Experiments}
\subsubsection{Dataset and Preprocessing}
The public SD-OCT dataset \cite{farsiu2014quantitative} includes both normal {\color{black}(265)} and age-related macular degeneration (AMD) {\color{black}(115)} cases.
The images were acquired using the Bioptigen Tabletop SD-OCT system (Research Triangle Park, NC).
The physical resolutions are 3.24 $\upmu$m (within A-scan), 6.7 $\upmu$m (cross A-scan), and 0.067 mm (cross B-scan).
Since the manual annotations are only available for a region centered at the fovea, subvolumes of size 400$\times$40$\times$512 ($N_A$, $N_B$, and $R$) voxels are extracted around the fovea.
We train the model on 263 subjects and test on the other 72 subjects {\color{black}(some cases are eliminated from analysis as the competing alignment algorithm \cite{pnevmatikakis2017normcorre} fails to handle them)}, which are randomly divided with the proportion of AMD cases unchanged.
The inner aspect of the inner limiting membrane (ILM), the inner aspect of the retinal pigment epithelium drusen complex (IRPE), and the outer aspect of Bruch's membrane (OBM) were manually traced.
For the multi-surface segmentation, there are two considerations.
First, we employ the topology guarantee module \cite{he2019fully} to make sure the correct order of the surfaces.
Second, the natural smoothness of these surfaces are different.
Therefore, we set different $\lambda$ (weight of $\mathcal{L}_\mathrm{SmoothS}$) values for different surfaces, according to the extents of smoothness and preliminary experimental results.
As to preprocessing, an intensity gradient method \cite{lang2013retinal} is employed to flatten the retinal B-Scan image to the estimated Bruch's membrane (BM), which can reduce memory usage.
When standalone B-scan alignment is needed, the NoRMCore algorithm \cite{pnevmatikakis2017normcorre} is employed.

For B-scan alignment, we adopt the mean absolute distance (MAD) of the same surface on two adjacent B-scans, and the average NCC between aligned B-scans for quantitative evaluation.
For retinal surface segmentation, the MAD between predicted and ground truth surface positions is used.
To compare the cross-B-scan continuity of the surfaces segmented by different methods, inspired by \cite{he2021structured}, we calculate the surface distance between adjacent B-Scans as the statistics of flatness and plot the histogram for inspection.

\subsubsection{Implementation}
The PyTorch framework (1.4.0) is used for experiments.
Implementation of our proposed network follows the architecture proposed in Model Genesis \cite{zhou2019models}, except that the 3D layers of the feature extractor $G_f$ is changed to 2D.
To reduce the number of network parameters, we halve the number of channels in each CNN block.
{\color{black}All networks are trained form scratch.}
Due to the memory limit, OCT volumes are cropped into patches of 320$\times$400$\times$40 voxels for training.
We utilize the Adam optimizer and train for 120 epochs.
The learning rate is initialized to 0.001 and halved when the {\color{black}loss} has not improved for ten consecutive epochs.
We train the network on three 2080 Ti GPUs with a mini-batch size of nine patches.
Based on preliminary experiments and natural smoothness of the three target surfaces, $\lambda$ is set to 0, 0.3, and 0.5 for ILM, IRPE, and OBM, respectively.
The source code is available at: https://github.com/ccarliu/Retinal-OCT-LayerSeg.git.

\begin{figure}[t]
\centering
\includegraphics[width=.55\textwidth,trim=0 20 0 20,clip]{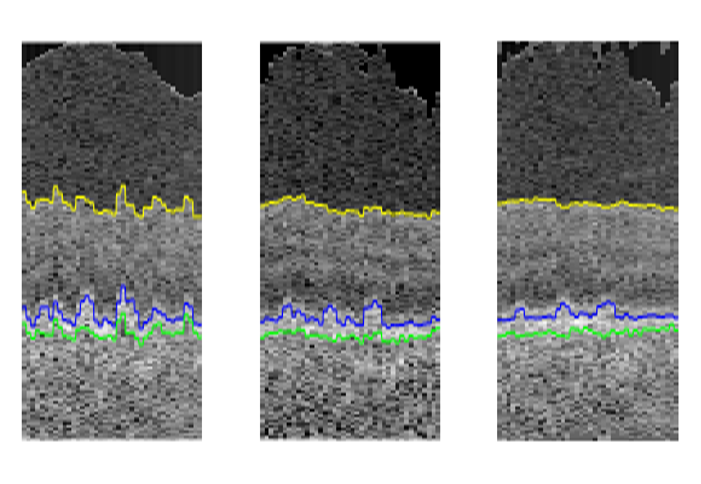}
\caption{B-scan alignment results visualized via cross sections.
Each B-Scan is repeated eight times for better visualization.
Left to right: no alignment (flattened to the BM)/NoRMCore \cite{pnevmatikakis2017normcorre}/ours.
Yellow: ILM, blue: IRPE, and green: OBM.} \label{alignment}
\end{figure}

\begin{table}[t]
	\caption{B-scan alignment results.
	MAD: mean absolute distance (in pixels).
	NCC: normalized cross-correlation.}\label{tab1}
    \centering
    \begin{adjustbox}{width=.8\textwidth}
	\begin{tabular}{l|ccccc}
		\hline
		Methods & ILM (MAD) & IRPE (MAD) & OBM (MAD) & Average (MAD) & NCC\\
		\hline
		No alignment & 3.91& 4.17& 3.93 & 4.00 & 0.0781\\
		NoRMCore \cite{pnevmatikakis2017normcorre} & 1.74& 2.19& 1.87 & 1.93 & 0.0818 \\
		Ours & \textbf{1.55}& \textbf{2.11}& \textbf{1.78} & \textbf{1.81} & \textbf{0.0894} \\
		\hline
	\end{tabular}
    \end{adjustbox}
	\label{result}
\end{table}

\subsubsection{B-Scan Alignment Results}
Figure \ref{alignment} shows the cross-B-scan sections of an OCT volume before and after alignment.
As we can see, obvious mismatches between B-scans can be observed before alignment, and both alignment algorithms make the B-scans more aligned.
While it is hard to tell from the visualizations, quantitative results in Table \ref{tab1} suggest that our framework aligns the B-scans better than the NoRMCore \cite{pnevmatikakis2017normcorre}, with generally lower MADs and higher NCCs.

\begin{table}[t]
	\caption{Mean absolute distance ($\upmu$m) as surface errors $\pm$ standard deviation.}\label{tab2}
	\centering
\begin{adjustbox}{width=0.8\textwidth}
	\begin{tabular}{l|cccccc}
		\hline
		Methods & FCBR \cite{he2019fully}  & Proposed & no\_align & pre\_align & no\_smooth & 3D-3D \\
		\hline
		ILM (AMD) & 1.73$\pm$2.50 & 1.76$\pm$2.39 & 2.25$\pm$3.77 & 1.80$\pm$2.36 & \textbf{1.68$\pm$1.84} & 1.87$\pm$2.19  \\
		
		ILM (Normal) & \textbf{1.24$\pm$0.51} & 1.26$\pm$0.47 & 1.40$\pm$0.42 & 1.30$\pm$0.49 & 1.27$\pm$0.47 & 1.31$\pm$0.46  \\
		
		IRPE (AMD) & 3.09$\pm$2.09 & \textbf{3.04$\pm$1.79} & 3.14$\pm$1.72 & 3.09$\pm$1.79 & 3.10$\pm$1.97& 3.12$\pm$1.74 \\
		
		IRPE (Normal) & 2.06$\pm$1.51 & 2.10$\pm$1.36 & 2.18$\pm$1.37 & \textbf{2.05$\pm$1.40} & 2.13$\pm$1.45 & 2.13$\pm$1.45 \\
		OBM (AMD) & 4.94$\pm$5.35 & \textbf{4.43$\pm$2.68} & 4.96$\pm$3.26 & 4.75$\pm$3.61 & 4.84$\pm$3.43 & 4.78$\pm$2.99 \\
		OBM (Normal) & \textbf{2.28$\pm$0.36} & 2.40$\pm$0.39 & 2.49$\pm$0.40 & 2.34$\pm$0.37 & 2.45$\pm$0.41 & 2.43$\pm$0.40 \\
		
		\hline
		
		Overall & 2.78$\pm$3.31 & \textbf{2.71$\pm$2.25} & 3.00$\pm$2.82 & 2.77$\pm$2.59 & 2.81$\pm$2.48 & 2.85$\pm$2.34\\
		
		\hline
	\end{tabular}
\end{adjustbox}
	\label{result}
\end{table}

\begin{figure}[t]
\centering
\includegraphics[width=.9\textwidth,trim=0 257 0 0,clip]{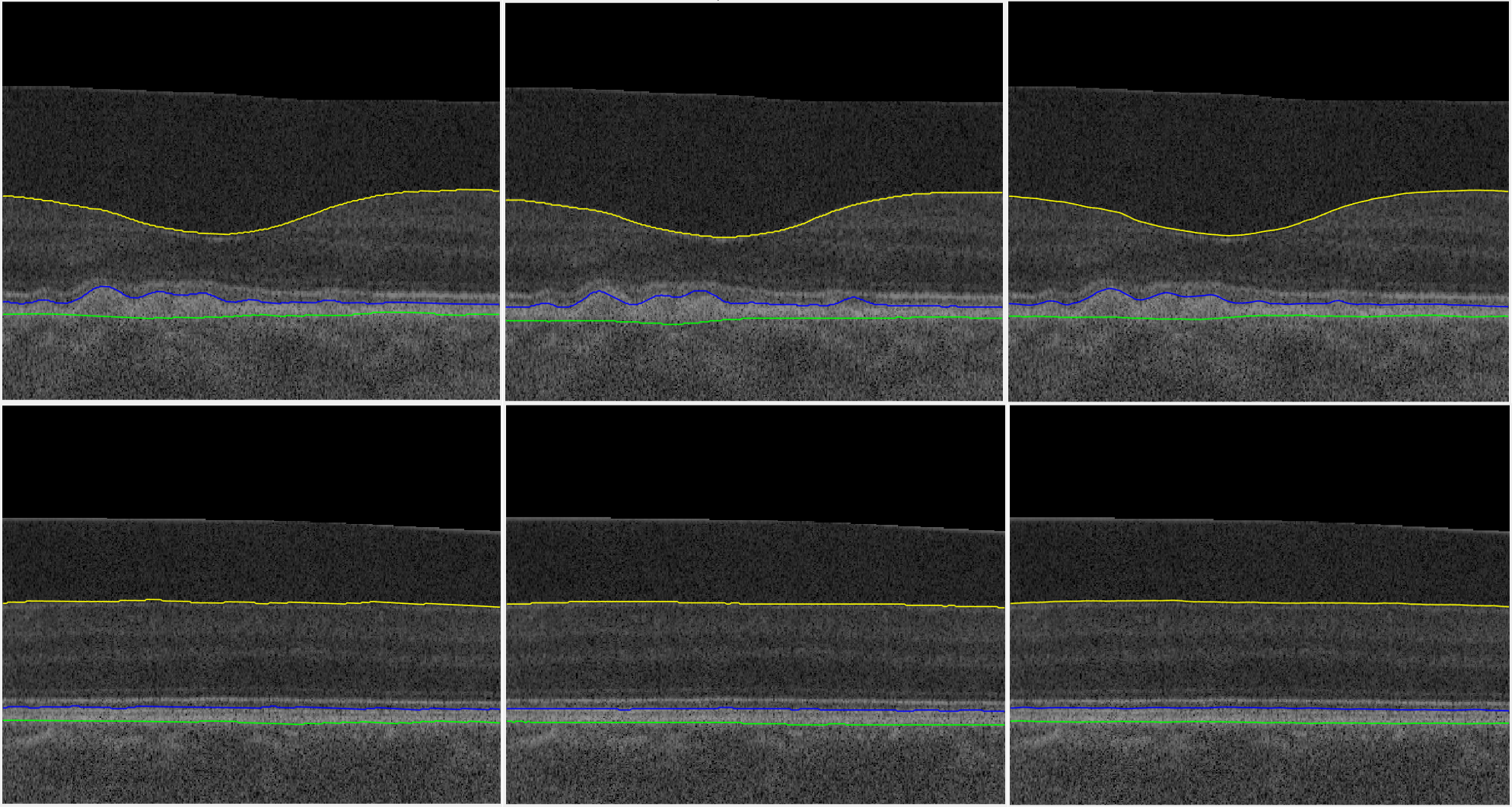}
\caption{Visualization of the manual segmentation (left), segmentation by FCBR \cite{he2019fully} (middle), and segmentation by our framework (right) of an AMD case.
Visualization of a normal control is shown in Fig. \ref{fig:sup_seg}.
Yellow: ILM, blue: IRPE, and green: OBM.} \label{fig:seg}
\end{figure}

\subsubsection{Surface Segmentation Results}
The results are presented in Table \ref{tab2}.
First, we compare our proposed method to FCBR \cite{he2019fully} (empirically tuned for optimal performance), which is a state-of-the-art method based on 2D surface regression.
As we can see, our method achieves lower average MADs with lower standard deviations (example segmentations in Figs. \ref{fig:seg} and \ref{fig:sup_seg}).
In addition, we visualize surface positions of BM as depth fields in Fig \ref{fig1}.
For a fair comparison, we visualize the FCBR results aligned by NoRMCore \cite{pnevmatikakis2017normcorre}.
It can be observed that our method (Fig. \ref{fig1}(d)) produces a smoother depth field than FCBR does (Fig. \ref{fig1}(c)).

\begin{figure}[t]
\centering
\includegraphics[width=.86\textwidth,,trim=15 13 15 7,clip]{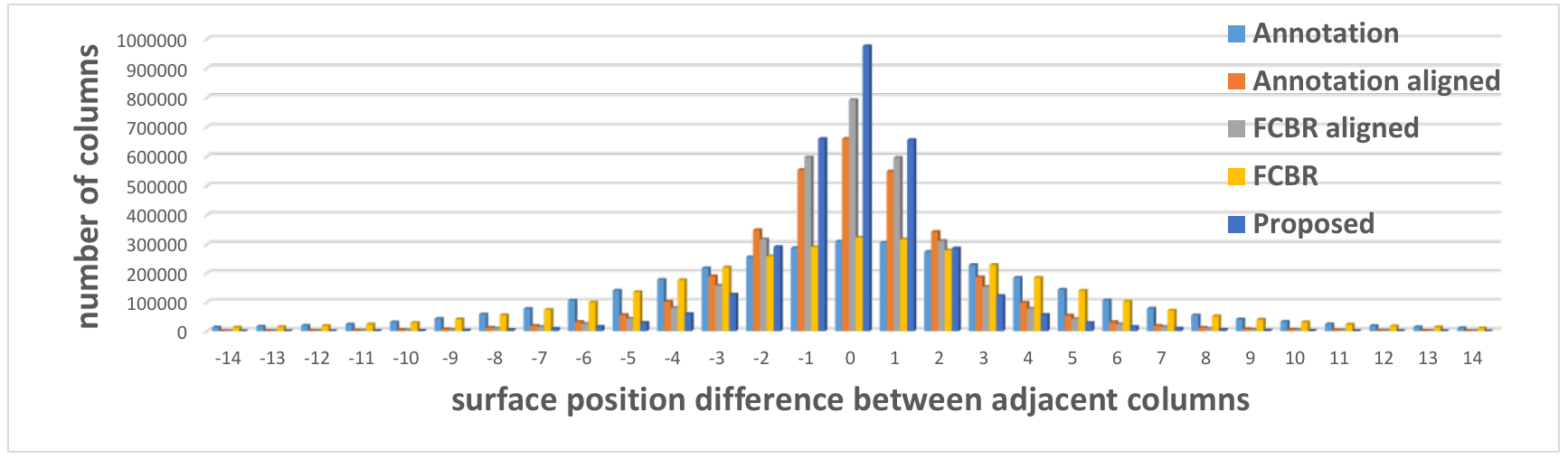}
\caption{Histogram of the surface distance (in pixels) between adjacent B-Scans.} \label{bscan_dis}
\end{figure}

Next, we conduct ablation experiments to verify the effectiveness of each module in the proposed framework.
Specifically, we evaluate several variants of our model: no\_smooth (without the global coherence loss $\mathcal{L}_\mathrm{SmoothS}$), no\_align (without the alignment branch or pre-alignment), pre\_align (without the alignment branch but pre-aligned by NoRMCore \cite{pnevmatikakis2017normcorre}), 3D-3D (replacing the encoder $G_f$ by 3D CNNs).
The results are presented in Table \ref{tab2}, from which several conclusions can be drawn.
First, the variant without any alignment yields the worst results, suggesting that the mismatch between B-scans does have a negative impact on 3D analysis of OCT data such as our 3D surface segmentation.
Second, our full model with the alignment branch improves over pre\_align.
We speculate this is because the alignment branch can produce better alignment results, and more importantly, it produces a slightly different alignment each time, serving as a kind of data and feature augmentation of enhanced diversity for the segmentation decoder $G_s$.
Third, removing $\mathcal{L}_\mathrm{SmoothS}$ apparently decreases the performance, demonstrating its effectiveness in exploiting the anatomical prior of smoothness.
Lastly, our hybrid 2D-3D framework outperforms its counterpart 3D-3D network, indicating that the 2D CNNs can better deal with the mismatched B-scans prior to proper realignment.

\subsubsection{B-Scans Connectivity Analysis}
As shown in Fig. \ref{bscan_dis}, surfaces segmented by our method has better cross-B-scan connectivity than those by FCBR \cite{he2019fully} even with pre-alignment, as indicated by the more conspicuous spikes clustered around 0.
This suggests that merely conducting 3D alignment does not guarantee 3D continuity, if the aligned B-scans are handled separately.
It is worth noting that our method achieves even better cross-B-scan connectivity than the manual annotations after alignment, likely due to the same reason (\emph{i.e.}, human annotators work with one B-scan at a time).

\section{Conclusion}
This work presented a novel hybrid 2D-3D framework for simultaneous B-scan alignment and retinal surface regression of volumetric OCT data.
The key idea behind our framework was the global coherence of the retinal layer surfaces both within and across B-scans.
Experimental results showed that our framework was superior to the existing state-of-the-art method \cite{he2019fully} for retinal layer segmentation, and verified the effectiveness of the newly proposed modules of our framework.
In the future, we plan to evaluate our framework on additional datasets with more severe diseases and more annotated layers.

\subsubsection{Acknowledgments.} This work was supported by the Fundamental Research Funds for the Central Universities (Grant No. 20720190012), Key-Area Research and Development Program of Guangdong Province, China (No. 2018B010111001), and Scientific and Technical Innovation 2030 - ``New Generation Artificial Intelligence'' Project (No. 2020AAA0104100).
%
%
%
\bibliographystyle{splncs04}
\bibliography{ref}
%
\newpage
\begin{center}
\textbf{\large Supplementary Material: Simultaneous Alignment and Surface Regression Using Hybrid 2D-3D Networks for 3D Coherent Layer Segmentation of Retina OCT Images}
\end{center}
\setcounter{equation}{0}
\setcounter{figure}{0}
\setcounter{table}{0}
\setcounter{page}{1}
\makeatletter
\renewcommand{\theequation}{S\arabic{equation}}
\renewcommand{\thefigure}{S\arabic{figure}}
\renewcommand{\thepage}{S\arabic{page}}
\begin{figure}[h]
\centering
\includegraphics[width=.9\textwidth,trim=0 0 0 256,clip]{segmentation_result_vis.png}
\caption{Visualization of the manual segmentation (left), segmentation by FCBR \cite{he2019fully} (middle), and segmentation by our framework (right) of a normal control.
Yellow: ILM, blue: IRPE, and green: OBM.} \label{fig:sup_seg}
\end{figure}

\begin{figure}[h]
\centering
\includegraphics[width=.95\textwidth,trim=0 10 0 5,clip]{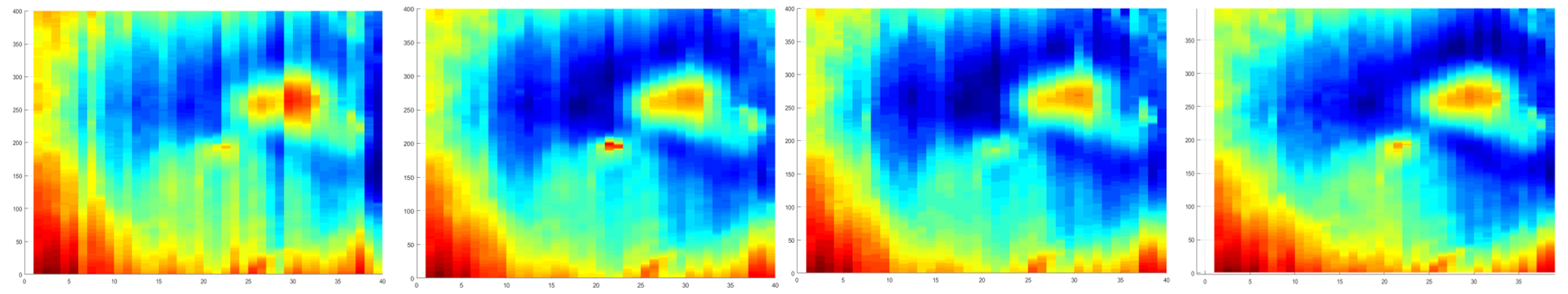}\\
\makebox[.75\textwidth][s]{\scriptsize(a)\hfill(b)\hfill(c)\hfill(d)}
\caption{Visualization of the BM surface as depth fields.
$x$-axis: cross-B-scan; $y$-axis: cross-A-scan.
(a) Our framework without the alignment branch.
(b) Our framework without the alignment branch but with the NoRMCore \cite{pnevmatikakis2017normcorre} pre-alignment.
(c)~FCBR~\cite{he2019fully} with NoRMCore \cite{pnevmatikakis2017normcorre} alignment. (d) Our full model.} \label{fig1}
\end{figure}
\end{document}